\documentclass[%
 reprint,
 superscriptaddress,
 showkeys,
 amsmath,amssymb,
 aps,
 pra,
 floatfix,
 twocolumn
]{revtex4-2}

\newcommand{\bra}[1]{\langle #1\rvert}
\newcommand{\ket}[1]{\lvert #1\rangle}

\usepackage{tabto}
\usepackage{graphicx,float}
\usepackage{dcolumn}
\usepackage{bm}
\usepackage{hyperref}
\usepackage{bigints}
\usepackage{amssymb}
\usepackage{accents}
\usepackage{tikz}
\usetikzlibrary{calc}
\usepackage{siunitx}
\AtBeginDocument{\RenewCommandCopy\qty\SI}

\usepackage{amsmath}
\usepackage{upgreek}
\usepackage{physics}
\usepackage{isomath}
\usepackage{placeins}
\usepackage{comment}
\usepackage{mathrsfs}

\begin{document}


\title{Coupled states of cold $\mathbf{^{174}Yb}$ atoms 
in a high-finesse cavity}
\author{Saran Shaju}
\affiliation{Experimentalphysik, Universität des Saarlandes, 66123 Saarbrücken, Germany}
\author{Dmitriy Sholokhov}
\affiliation{Experimentalphysik, Universität des Saarlandes, 66123 Saarbrücken, Germany}
\author{Simon B. J\"ager}
\affiliation{Physics Department and Research Center OPTIMAS, University of Kaiserslautern-Landau, 67663 Kaiserslautern, Germany}
\author{J\"urgen Eschner}
\affiliation{Experimentalphysik, Universität des Saarlandes, 66123 Saarbrücken, Germany}
\date{\today}


\begin{abstract}
We experimentally and theoretically study the formation of dressed states emerging from strong collective coupling of the narrow intercombination line of Yb atoms to a single mode of a high-finesse optical cavity. By permanently trapping and cooling the Yb atoms during their interaction with the cavity, we gain continuous experimental access to the dressed states. This allows us to detect both their field and their atomic properties, by simultaneously measuring the steady-state cavity transmission and free-space fluorescence. By varying the cavity and probe frequencies, we observe coupled atom-cavity states with atom number-dependent splitting, the hallmark of collective strong coupling of the atoms with the single cavity mode. We find additional fluorescence output at atomic resonance, which we explain by the effects of dephasing and inhomogeneous broadening. We compare our experimental results with a theoretical model and find good qualitative agreement.
\end{abstract}


\maketitle

\section{I. Introduction}
Cold atomic ensembles in optical cavities are a versatile platform to study collective effects such as self-organization~\cite{Domokos:2002,Black:2003,Asboth:2005,Arnold:2012,Ritsch:2013,Schuetz:2015}, super- and subradiance~\cite{Gross:1982,Baumann:2010,Guerin:2016,Gegg:2018,Norcia:2016,Norcia:2018,Klinder:2015,Laske:2019,Shankar:2021}, and in general long-range or even arbitrary cavity-mediated interactions~\cite{Norcia:2018:2,Marsh:2021,Kroeze:2023,Marsh:2024,Defenu:2023}. Based on these effects various applications have been proposed reaching from quantum simulators and computers~\cite{Torggler:2017,Altman:2021,Ye:2023} to ultra-precise lasers and active optical clocks~\cite{Chen:2009,Meiser:2009,Liu:2020,Kristensen:2023,Bohr:2024}. 
Trapping light in the confined geometry of an optical cavity massively enhances the interaction between a photon and a single atom. 
In the presence of many atoms, this interaction is in addition collectively enhanced which allows for very strong light-matter interactions and the achievement of very large collective linewidths even when working with atoms that possess very narrow natural linewidths. With this one can work in a regime where very strong non-linearities can be generated while unwanted decoherence channels such as spontaneous emission are weak which is an ideal ground for future quantum technologies. 
	
However, this field faces a major challenge which arises from atomic motion inducing line-broadening and also eventually the loss of atoms. Cooling and trapping~\cite{metcalf:1999} of the atoms in a regime where the total broadening is smaller than the collectively enhanced coupling strength is a minimum requirement. In several experiments this is achieved in a sequential way where first the gas is cooled and prepared in a trap and then one investigates collective effects emerging from the atom-cavity coupling in a second step. In this second step the atoms will heat which leads to the loss of atoms and sets the time limit on which operations and simulations can be performed. This is detrimental and only recently several groups in this field aim at operating such devices in a continuous regime where atoms are constantly refilled~\cite{Chen:2019,Tang:2022,Liu:2020,Jaeger:2021,Cline:2022,Chen:2022} or where the atom-cavity interactions operate simultaneously with trapping and cooling~\cite{Salzburger:2004,Salzburger:2006,Xu:2016,Jaeger:2017,Gothe:2019,Gothe:2019:2,Hotter:2019}.
	
The latter is the situation that we investigate in this paper. We experimentally and theoretically investigate the formation of dressed states formed by strong collective coupling between the $^1$S$_0\,\leftrightarrow\,^3$P$_1$ intercombination line of $^{174}$Yb atoms and a resonant optical cavity. Formation of dressed states has been studied in several works reaching from single-atom experiments~\cite{Agarwal:85,Zhu:1990,Thompson:1992,haroche:1996,Kimble:1998,McKeever:2004,Schuster:2008,Mücke:2010} to atomic ensembles~\cite{Gripp:1997,McKeever:2004,Tuchman:2006,Suarez:2023}. More recently these studies have also been extended to narrow intercombination lines~\cite{Christensen:2015,Cline:2022,Rivero:2023}. In this work we investigate a similar situation but observe both the atomic and cavity degrees while operating a magneto-optical trap (MOT) on the $^1$S$_0\,\leftrightarrow\,^1$P$_1$ transition which permanently traps and cools the atoms. This allows us to study their dynamics over long timescales that are of the orders of several seconds. Over such long timescales we detect the field as well as the atomic character of the dressed states by measuring the cavity transmission and by detecting the fluorescence from the narrow intercombination line, respectively. 
We compare these measurements to theoretical predictions that include inhomogeneous broadening of the narrow atomic line. We find qualitative good agreement for reasonable atom numbers and broadening. Our work is a step towards exploring the behavior of narrow-linewidth atoms strongly coupled to optical cavities, while they are continuously trapped on a broader line, which a possible scenario for their use in quantum technologies.

This paper is structured as follows. In Sec.~\ref{Sec:2} we describe the experimental setup while Sec.~\ref{Sec:3} introduces our theoretical tools. In Sec.~\ref{Sec:4} we report the measurements of the cavity field transmission and the free-space fluorescence and compare them to theory. In Sec.~\ref{Sec:5} we summarize our results and conclude.

\section{Experimental Setup and Measurement Protocol\label{Sec:2}}

\subsection{Lasers}\label{Sec:2a}

Two lasers are employed. One is used for cooling and trapping the Yb atoms in a conventional 6-beam magneto-optical trap (MOT) on the $^1$S$_0\,\leftrightarrow\,^1$P$_1$ dipole-allowed transition at $\lambda_{\mathrm{MOT}}=\qty{399}{nm}$ ("blue", linewidth $\gamma_{\mathrm{MOT}}=2\pi\times\qty{29.1}{MHz}$). The other laser is near-resonant with the $^1$S$_0\,\leftrightarrow\,^3$P$_1$ intercombination line at $\lambda=\qty{556}{nm}$ ("green", linewidth $\gamma=2\pi\times\qty{182.4}{kHz}$), it is used as probe to drive the atom-cavity system through a cavity mirror. The blue laser is frequency-stabilized via a transfer locking scheme \cite{Rohde:2010}, the green laser is stabilized to the green transition by atomic-beam spectroscopy. Both lasers are tuned with acousto-optical modulators (AOMs). The relevant transitions are displayed in Fig.~\ref{Fig:1}(b).

\subsection{Atoms}\label{Sec:2b}

\begin{figure}[ht]
\includegraphics[width=1\columnwidth]{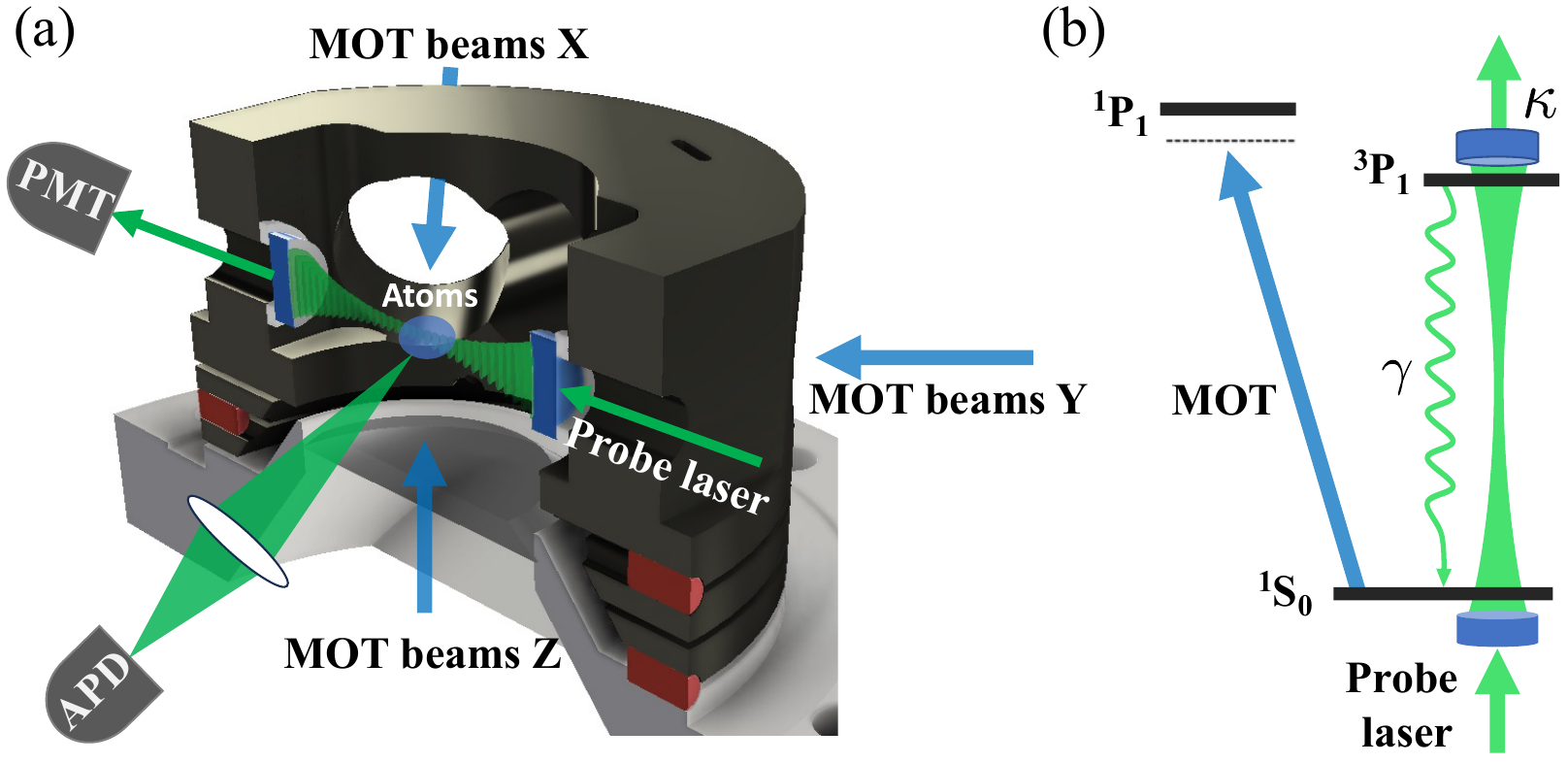}
  \caption{(a) Schematic of the experimental setup for probing the atom-cavity coupled states. Atoms are held by a conventional 3D MOT on the $^1$S$_0\,\leftrightarrow\,^1$P$_1$ blue line, and a laser drive is provided along the cavity axis. The cavity transmission is detected using a photo-multiplier tube (PMT) and an avalanche photo detector (APD) is used for counting the fluorescence photons emitted by the atoms into free space. (b) The atomic level scheme shows the relevant transitions for the experiment, where the atoms are held using the broad blue transition and the atom-cavity interaction is on the narrow $^1$S$_0\,\leftrightarrow\,^3$P$_1$ intercombination line. }
  \label{Fig:1}
\end{figure}

In the experiment we work with $^{174}$Yb atoms that are evaporated from an oven operating at $\qty{500}{\degreeCelsius}$. The collimated atoms are guided to a magneto-optical trap (MOT) and slowed down by a Zeeman slower operating on the $^1$S$_0\,\leftrightarrow\,^1$P$_1$ transition. The MOT, typically operating at $\Delta_{\text{MOT}}=-2\pi\times\qty{30}{MHz}$, traps and cools the atoms in a $\sim\qty{1}{mm}$-size spatial region around the center of our optical cavity, see Fig.~\ref{Fig:1}(a). In the overlap region between MOT and cavity mode, about $10^4$ to $10^5$ atoms are held at 5 to $\qty{10}{mK}$ temperature.

\subsection{Cavity}\label{Sec:2c}

The high-finesse optical cavity has a length of $L=\qty{4,78}{cm}$ in Fabry-Perot configuration and $\qty{87}{\micro m}$ waist radius of the TEM$_{00}$ mode. It couples to the $^1$S$_0\,\leftrightarrow\,^3$P$_1$ atomic transition at $\lambda=\qty{556}{nm}$, where it has a loss rate of $\kappa=2\pi\times\qty{70}{kHz}$ (finesse $\mathcal{F}=$~\qty{45000}). A single atom in this setup couples to the cavity with a vacuum Rabi frequency
$g_0=2\pi\times\qty{66}{kHz}$, corresponding to a single-atom cooperativity $C_1=g_0^2/(\kappa\gamma)=0.34$ \cite{Suzuki:2011}. By trapping a large number $N$ of atoms in the cavity mode, collective strong coupling $C = N C_{1} \gg 1$ is achieved, which leads to the formation of atom-cavity coupled states in the system that we investigate. Nevertheless, since the beam waist of the cavity is small compared to the size of the cloud itself, only a small percentage of the total number of atoms interact with the cavity.

The cavity length is stabilized by locking it to a reference derived from the green laser. In our setup, we use a reference beam which is $\qty{0.4}{GHz}$ shifted from the green transition frequency and resonant to the $15^\text{th}$ transverse mode of the cavity. Feedback is applied to piezo-electric actuators attached to the cavity mirrors. The uncertainty of the cavity frequency is on the order of the cavity linewidth. The cavity frequency is then controlled by shifting the frequency of the reference beam with an AOM.

\subsection{Detection}\label{Sec:2d}

Detection in our setup happens in two channels: first, detection of the cavity transmission, and second, detection of the atomic fluorescence emitted into free space. The transmitted light from the cavity is collected to a single mode fiber and detected with a photo-multiplier tube (PMT). In the figures shown below, 1 mV of PMT signal translates to about \qty{70}{pW} of cavity output, or $\sim 4 \times 10^{4}$ photons inside the cavity. Fluorescence is recorded using a telescope imaging scheme with a pair of 1-inch lenses to collect and collimate the photons, which are then fiber-coupled to an avalanche photo detector (APD) in counting mode. A color filter rejects the blue MOT fluorescence. 

\subsection{Measurement protocol}\label{Sec:2e}

The measurement begins with loading the atoms into the MOT, tracking the fluorescence at $\qty{399}{nm}$ as a measure of number of atoms trapped. The atomic flux to the MOT is regulated using a motorized mechanical shutter. Using this we control the number of atoms that interact with the cavity. The cloud position is adjusted by the MOT beam directions and powers, and monitored with CCD cameras to ensure the geometrical overlap with the cavity mode. Once good overlap is attained that supports the formation of the atom-cavity states, the cavity is driven with the probe laser. 

With the atoms inside the cavity, we set the detuning of the cavity to a fixed value, $\Delta_\text{c}$, and scan the frequency of the probe laser, $\Delta_\text{p}$. For each parameter set ($\Delta_\text{c}, \Delta_\text{p}$) we measure the cavity transmission and free-space fluorescence using the PMT and APD, respectively, for 5 to 10 seconds to obtain a good signal-to-noise ratio. 

We emphasize some aspects in which our experimental situation is peculiar, and regarding which our results are novel. First, the atom-cavity states are subject to dephasing induced by (i) the MOT beams and (ii) by the strong saturation of the green transition inside the cavity. Second, on the scale of the cavity and atomic linewidth, there is significant inhomogeneous (Doppler) broadening due to the MOT temperature. To overcome these effects we trap a large number of atoms and produce a Rabi splitting nearly 2 orders of magnitude larger than the linewidth. Finally, by measuring the fluorescence when the strong probe is inside the cavity, we do not only observe the cavity transmission but also have simultaneous access to the atomic response of the coupled system.

\section{Theoretical description\label{Sec:3}}

\subsection{Mean-field description}
We use a mean-field description for the dynamics of $N$ two-level atoms with internal ground (excited) state $\ket{g}$ ($\ket{e}$) interacting with the driven cavity mode. This means we replace all quantum operators by their expectation values. The equations of motion describing the coupled atom-cavity dynamics are given by
	\begin{align}
		\frac{d\alpha}{dt}=&\left[i(\Delta_p-\Delta_c)-\frac{\kappa}{2}\right]\alpha-i\frac{\eta}{2}-i\frac{g_0}{2}\sum_js_j\label{eq:alpha}\\
   	\frac{ds_j}{dt}=&\left[i(\Delta_p-\omega_j)-\frac{\gamma+\gamma_d}{2}\right]s_j+i\frac{g_0}{2}\alpha z_j\label{eq:s}\\
		\frac{dz_j}{dt}=&-\gamma(1+z_j)+ig_0(\alpha^*s_j-s_j^*\alpha)\label{eq:z},
	\end{align}
where $\alpha=\langle\hat{a}\rangle$, $s_j=\langle\hat{\sigma}_j^-\rangle$, and $z_j=\langle\hat{\sigma}_j^{z}\rangle$ are the mean values of the cavity field annihilation operator $\hat{a}$, the lowering operator $\hat{\sigma}_j^-=\ket{g}_j\bra{e}$, and the inversion operator $\hat{\sigma}_j^z=\ket{e}_j\bra{e}-\ket{g}_j\bra{g}$, respectively. This description is derived in a frame rotating with the laser frequency $\omega_L$, and we have introduced the respective detunings between cavity and atom, $\Delta_c=\omega_c-\omega_a$, and between laser and atom, $\Delta_p=\omega_p-\omega_a$. In Eq.~\eqref{eq:alpha} we have introduced the cavity linewidth $\kappa$, the cavity-field driving amplitude $\eta$, and the vacuum Rabi frequency $g_0$. The frequencies governing the dynamics of the atomic degrees of freedom, Eqs.~\eqref{eq:s}-\eqref{eq:z}, are the natural atomic linewidth $\gamma$, homogeneous dephasing $\gamma_d$, and an atom-dependent frequency $\omega_j$ which is drawn from a distribution function
	\begin{align}
		f(\omega)=\frac{1}{\sqrt{2\pi\Delta\omega^2}}e^{-\frac{\omega^2}{2\Delta\omega^2}}.\label{eq:f}
	\end{align}
Inhomogeneous broadening described by $\Delta\omega$ and homogeneous dephasing described by $\gamma_{\text{d}}$ is introduced to model several effects in our experiment including Doppler broadening, magnetic shifts, and random motion introduced by the MOT. 

Equations~\eqref{eq:alpha}-\eqref{eq:z} are the basis of our theoretical analysis, and in the following we will derive the stationary state which is used for comparison to the experimental results.

\subsection{Stationary state}
 
We will now derive the stationary state of Eqs.~\eqref{eq:alpha}-\eqref{eq:z}. We are interested in both the cavity field and in the excited state population, which are accessed in the experiment by measuring the cavity transmission and the atomic fluorescence, respectively.
	
We find the stationary solution of Eq.~\eqref{eq:s}, which reads 
	\begin{align}
		s_j=\frac{-\frac{g_0}{2}z_j}{(\Delta_p-\omega_j)+i\frac{\gamma+\gamma_d}{2}}\alpha\label{eq:sts}.
	\end{align}
Using this result in Eq.~\eqref{eq:z} and imposing $dz_j/dt=0$, we obtain the steady state expression for $z_j$ which takes the form
	\begin{align}
		z_j=-\frac{1}{1+\frac{\frac{g_0^2|\alpha|^2(\gamma+\gamma_d)}{2\gamma}}{(\Delta_p-\omega_j)^2+\left(\frac{\gamma+\gamma_d}{2}\right)^2}}.
	\end{align}
With this solution we find $s_j$ [Eq.~\eqref{eq:sts}] which is then used to solve $d\alpha/(dt)=0$ in Eq.~\eqref{eq:alpha}, such that 
	\begin{align}
		\alpha=\frac{\frac{\eta}{2}}{(\Delta_p-\Delta_c)+i\frac{\kappa}{2}+i\frac{\Gamma}{2}}.\label{eq:stalpha}
	\end{align}
Here, we introduced
	\begin{align}
		\Gamma = i& \sum_j\frac{\frac{\frac{g_0^2}{2}}{(\Delta_p-\omega_j)+i\frac{\gamma+\gamma_d}{2}}}{1+\frac{\frac{g^2_0|\alpha|^2(\gamma+\gamma_d)}{2\gamma}}{(\Delta_p-\omega_j)^2+\left(\frac{\gamma+\gamma_d}{2}\right)^2}}\nonumber\\
		= i& \int_{-\infty}^{\infty} d\omega\,F(\omega)\frac{\frac{\frac{g_0^2}{2}}{(\Delta_p-\omega)+i\frac{\gamma+\gamma_d}{2}}}{1+\frac{\frac{g_0^2|\alpha|^2(\gamma+\gamma_d)}{2\gamma}}{(\Delta_p-\omega)^2+\left(\frac{\gamma+\gamma_d}{2}\right)^2}}\nonumber
	\end{align}
using the function
	\begin{align}
		\tilde{f}(\omega)=\sum_j\delta(\omega-\omega_j).
	\end{align}
For very large atom numbers $N$ this function approaches $\tilde{f}(\omega)\to N f(\omega)$ with the distribution function introduced in Eq.~\eqref{eq:f}. We get then
\begin{align}
    \Gamma=i\int_{-\infty}^{\infty} d\omega\,f(\omega)\frac{\frac{\frac{Ng_0^2}{2}}{(\Delta_p-\omega)+i\frac{\gamma+\gamma_d}{2}}}{1+\frac{\frac{g^2_0|\alpha|^2(\gamma+\gamma_d)}{2\gamma}}{(\Delta_p-\omega)^2+\left(\frac{\gamma+\gamma_d}{2}\right)^2}}.\label{eq:Gamma}
\end{align}
Note that for $\Delta_\text{p}=0=\Delta\omega$ and zero dephasing $\gamma_d=0$ we have $\Gamma=NC\kappa$, with the cooperativity $C=g_0^2/(\kappa\gamma)$. More generally, $\Gamma$ represents the line broadening and dispersive shift of the cavity due the the presence of the atoms.

At last we calculate the average photon number leaving the cavity per unit time
    \begin{align}
    	T=\kappa|\alpha|^2=\frac{\frac{\eta^2}{4}}{\left|\Delta_p-\Delta_c+i\frac{\kappa}{2}+i\frac{\Gamma}{2}\right|^2}\label{eq:T}
    \end{align}
and fluorescent photon number per unit time
    \begin{align}
    F=\gamma\sum_j\frac{1+z_j}{2}=\int_{-\infty}^{\infty} d\omega\frac{\frac{\frac{g_0^2|\alpha|^2(\gamma+\gamma_d)}{4}}{(\Delta_p-\omega)^2+\left(\frac{\gamma+\gamma_d}{2}\right)^2}}{1+\frac{\frac{g_0^2|\alpha|^2(\gamma+\gamma_d)}{2\gamma}}{(\Delta_p-\omega)^2+\left(\frac{\gamma+\gamma_d}{2}\right)^2}}.\label{eq:F}
    \end{align}
We solve self-consistently Eq.~\eqref{eq:Gamma} and Eq.~\eqref{eq:T} for $|\alpha|^2$ which is then used to calculate the expressions $T$ and $F$. In the figures shown later we normalize the quantities $T$ and $F$ by the empty cavity transmission $T_0=\eta^2/\kappa$ and the collective spontaneous decay rate $N\gamma$.

In the next section we use the results of this model and compare them to our experimental data. 

\section{Collective light-matter states\label{Sec:4}}

In this section we present the main results which include the measured cavity transmission and fluorescence and its comparison to the theory described in the previous section.

\subsection{Measured cavity transmission and fluorescence}

\begin{figure}[ht]
    \includegraphics[width=1\columnwidth]{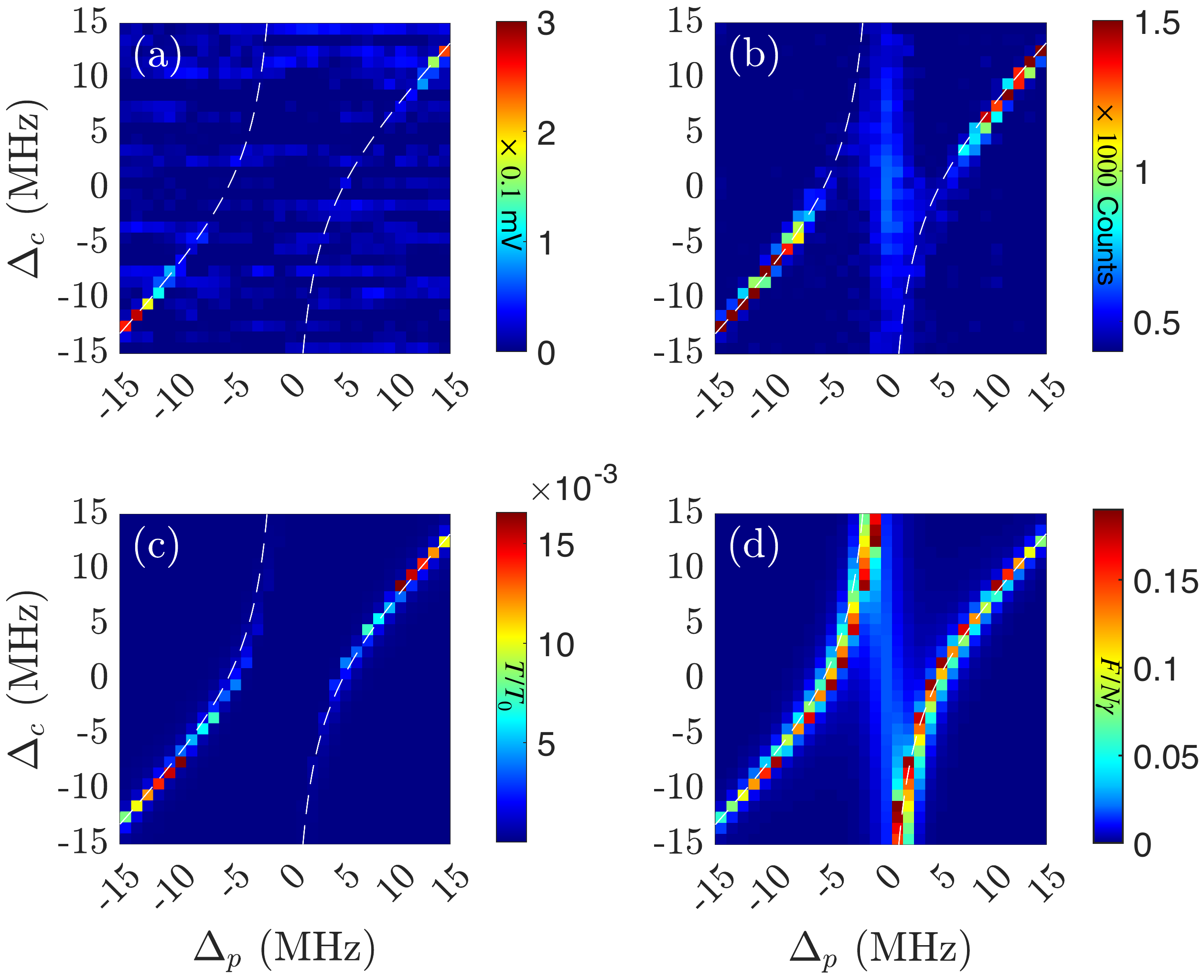}
    \caption{\label{Fig:2} Atom-cavity coupled states. Measurement (top row) and theory (bottom row) for cavity transmission (left) and fluorescence (right). (a) The cavity transmission detected with a PMT 
    when an effective atom number of 25,000 is present in the cavity mode. The white dashed line marks the resonance position according to the simple dressed-state formula, Eq.~\eqref{eq:sqrtN}. A background originating mainly from detector noise has been subtracted. (b) Simultaneously measured atomic fluorescence photon counts in 10 seconds. Atom-cavity coupled states are observed as in the case of cavity transmission. The peculiar feature which appears to be an additional broad resonance is visible around $\Delta_\text{p} \approx 0$. (c) and (d) show the calculated normalized cavity transmission $T/T_0$, Eq.~\eqref{eq:T}, and fluorescence $F/(N\gamma)$, Eq.~\eqref{eq:F}, respectively, when the cavity is driven with $\eta=2\pi\times\qty{87}{MHz}$ and an inhomogeneous broadening of $\Delta\omega=2\pi\times\qty{0.9}{MHz}$ and dephasing of $\gamma_\text{d}=2\pi\times\qty{1}{kHz}$ 
are used.}
\end{figure}      

In Fig.~\ref{Fig:2} we show (a) the measured cavity transmission in units of the voltage detected by the PMT and (b) the measured fluorescence in units of photon counts measured by the APD, as functions of the cavity-atom detuning $\Delta_\text{c}$ and probe-atom detuning $\Delta_\text{p}$. Both detunings are varied in a range $2\pi\times (\qty{-15}{MHz} \dots \qty{15}{MHz})$. In the cavity signal, Fig.~\ref{Fig:2}(a), we observe transmission for large detunings $|\Delta_\text{c}|, |\Delta_\text{p}| \ge 2\pi\times\qty{10}{MHz}$, and when $\Delta_\text{c} \approx \Delta_\text{p}$. Closer to atomic resonance, the transmission peaks show an avoided crossing and their magnitude tends to zero. The fluorescence signal, displayed in Fig.~\ref{Fig:2}(b), shows a similar behavior as the transmission, but additionally it exhibits a bright central feature close to resonance $\Delta_\text{c} \approx 0$ across nearly the whole range of $\Delta_\text{p}$. This feature does not appear in the cavity transmission. To understand this behavior better we apply our theoretical model and show the corresponding numerical results in Fig.~\ref{Fig:2}(c) and (d). The cavity transmission displayed in Fig.~\eqref{Fig:2}(c) is in good agreement with the experimental data in Fig.~\ref{Fig:2}(a). The physical origin of the avoided crossing is the formation of coupled eigenstates between the atoms and the cavity. The coupled state resonances are approximately described by 
    \begin{align}\label{eq:sqrtN}
    \Delta'_{\text{c}\pm} = \frac{\Delta_\text{c}}{2}\pm\frac{1}{2}\sqrt{\Delta_\text{c}^2+Ng_0^2}.
    \end{align}
We show the result of this formula as a guide to the eye in Fig.~\ref{Fig:2}(a)-(d) for $N=$~\qty{25000}, which is thereby used to estimate the effective number of atoms participating in the atom-cavity interaction.

The coupled states are also visible in the fluorescence, Fig.~\ref{Fig:2}(b) and (d), as their atomic contribution undergoes spontaneous decay into free space. Importantly, in the theory prediction in Fig.~\ref{Fig:2}(d) we also find the central feature close to atom-probe resonance, $\Delta_\text{p} \approx 0$, which is not part of the formation of atom-cavity coupled states. It will be investigated in more detail in section~\ref{sec:Fl_center}.

\subsection{Atom number dependence}\label{sec:Fl_vs_N}

\begin{figure}[t]
    \includegraphics[width=1\columnwidth]{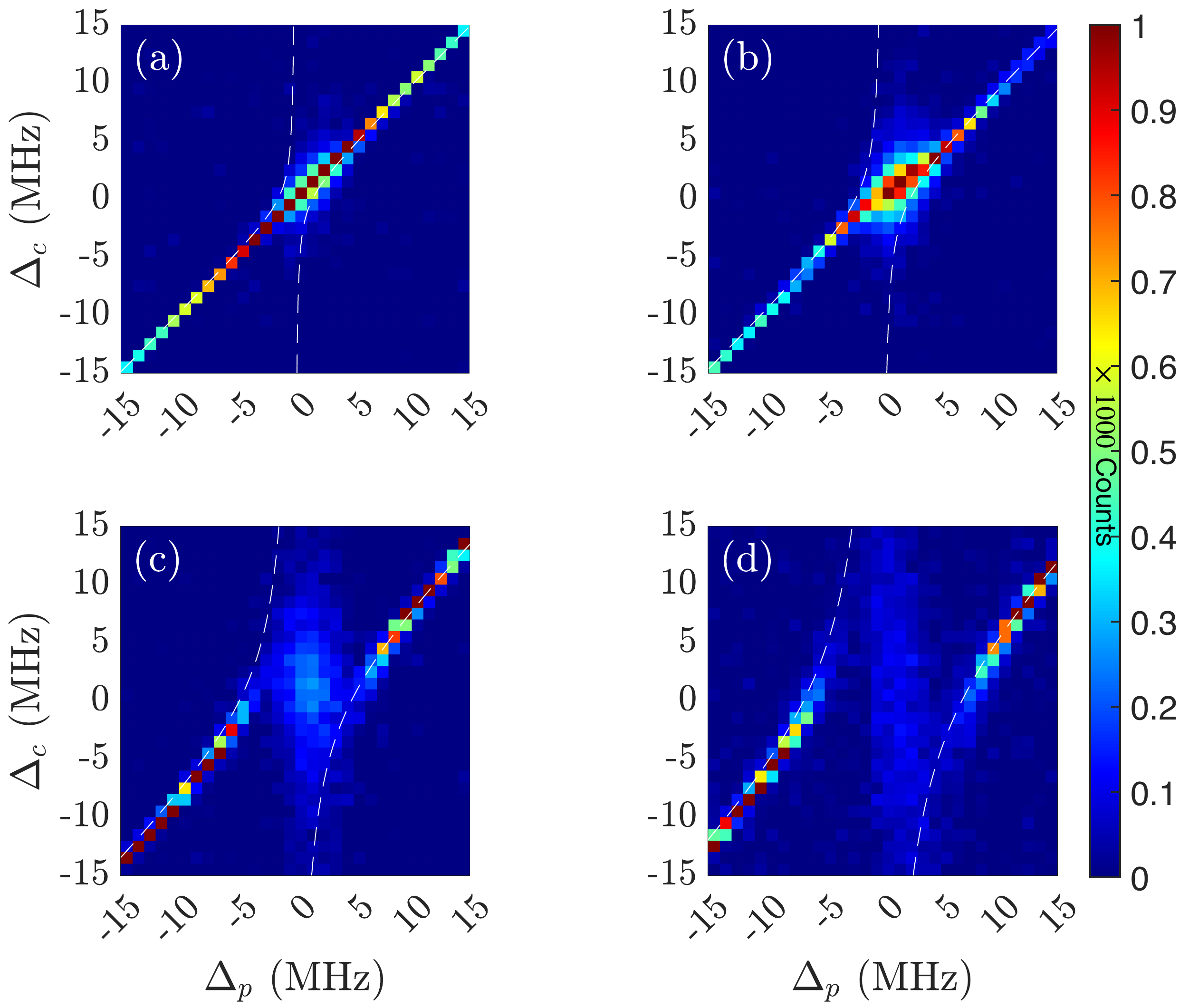}
    \caption{\label{Fig:Fl_vs_N} Fluorescence from coupled atom-cavity states as function of atom number. In (a) to (d), the  effective number of atoms interacting with the cavity mode is 
    $N =$ \qty{2000}, \qty{5000}, \qty{21250}, \qty{42500}, 
    respectively, according to the fitted white-dashed line. Fluorescence is counted for 10 seconds per pixel of the maps; background from APD dark counts has been subtracted.
    } 
\end{figure}

Controlling the atomic flux into the trapping volume allows us to extend our experiments and include variation of the atom number. We focus here on showing the fluorescence excited via the probe, measured as in the previous section, but for different numbers of atoms in the cavity mode. Figure~\ref{Fig:Fl_vs_N} shows the results when $\Delta_\text{c}$ and $\Delta_\text{p}$ are varied in the same range as in Fig.~\ref{Fig:2}. As we go from Fig.~\ref{Fig:Fl_vs_N}(a) to (d), the atom number is increased and correspondingly more splitting is visible. In Fig.~\ref{Fig:Fl_vs_N}(a), corresponding to the lowest number of atoms, the fluorescence appears very close to the diagonal, $\Delta_\text{c} = \Delta_\text{p}$; for the next higher atom number one can observe the beginning of splitting close to the atomic resonance.
As we continue to increase the atom number, in Fig.~\ref{Fig:Fl_vs_N}(c) a clear splitting is observed, that becomes noticeably wider as we go to Fig.~\ref{Fig:Fl_vs_N}(d). In the latter two maps, the splitting is accompanied by the already observed central fluorescence maximum; for the smaller values of atom number, that feature masks the splitting to some extent. 

The normal mode splitting predicted by Eq.~\eqref{eq:sqrtN} is shown in Fig.~\ref{Fig:Fl_vs_N}(a)-(d) as dashed white lines. The observations indicate that for the range of atom numbers that we are able to explore, we reach far into the collective strong coupling regime, where the separation of the coupled resonances of the atom-cavity system is larger than the characteristic frequencies of all relevant broadening mechanisms.

\subsection{Central feature in the fluorescence}\label{sec:Fl_center}

\begin{figure}[ht]
    \includegraphics[width=1\columnwidth]{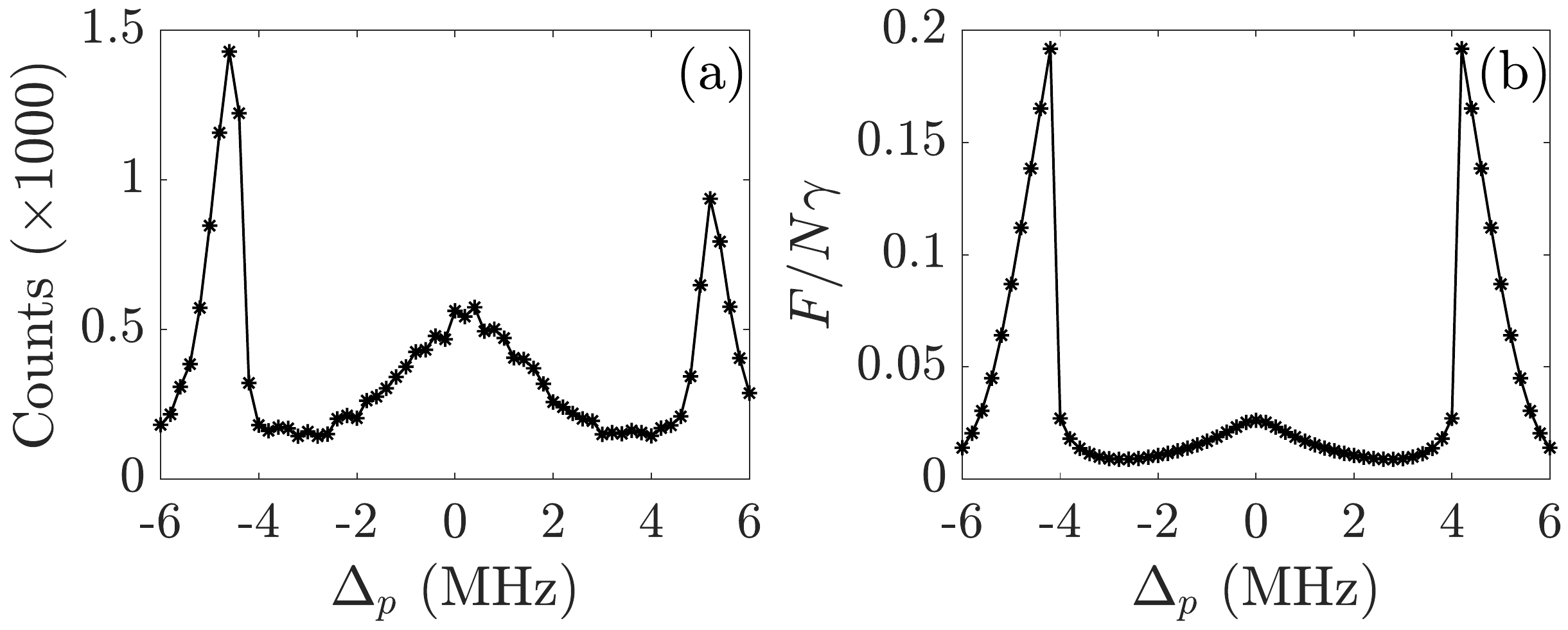}
    \caption{\label{Fig:Central} Central fluorescence maximum. Atomic fluorescence as a function of laser detuning when the cavity is resonant to the atomic transition, $\Delta_\text{c} = 0$. (a) Measured photon counts using an APD vs. probe laser frequency, with \qty{200}{kHz} resolution. (b) Calculated fluorescence using Eq.~\eqref{eq:F}, where an effective atom number of \qty{22500} is used with a cavity drive strength of $\eta=2\pi\times\qty{80}{MHz}$ and inhomogeneous broadening of $\Delta\omega=2\pi\times\qty{0.62}{MHz}$.}
\end{figure}

In this section we investigate the fluorescence from the atoms with higher frequency resolution, in particular the central feature which, conspicuously, does not appear in the cavity transmission. In Fig.~\ref{Fig:Central}(a), the detected fluorescence is plotted as a function of probe-atom detuning when the cavity-atom detuning is set to zero. The laser frequency is scanned over the range $\Delta_\text{p} = 2\pi\times (\qty{-6}{MHz} \dots \qty{6}{MHz})$ in steps of $2\pi\times\qty{200}{kHz}$, and for each laser frequency the fluorescence photons are counted for 5 seconds. The measurement shows two peaks around $\pm\qty{5}{MHz}$ and a broad peak centered around atomic resonance. The side peaks are understood as the atom-cavity coupled states, and their separation is described by $g_0\sqrt{N}$ (Eq.~\eqref{eq:sqrtN} with $\Delta_\text{c}=0$) corresponding to an effective atom number of \qty{22500}. Our prediction from theory calculated using Eq.~\eqref{eq:F} is displayed in Fig.~\ref{Fig:Central}(b) and shows reasonable agreement with the experimental observation. 

\begin{figure}[t]
    \includegraphics[width=1\columnwidth]{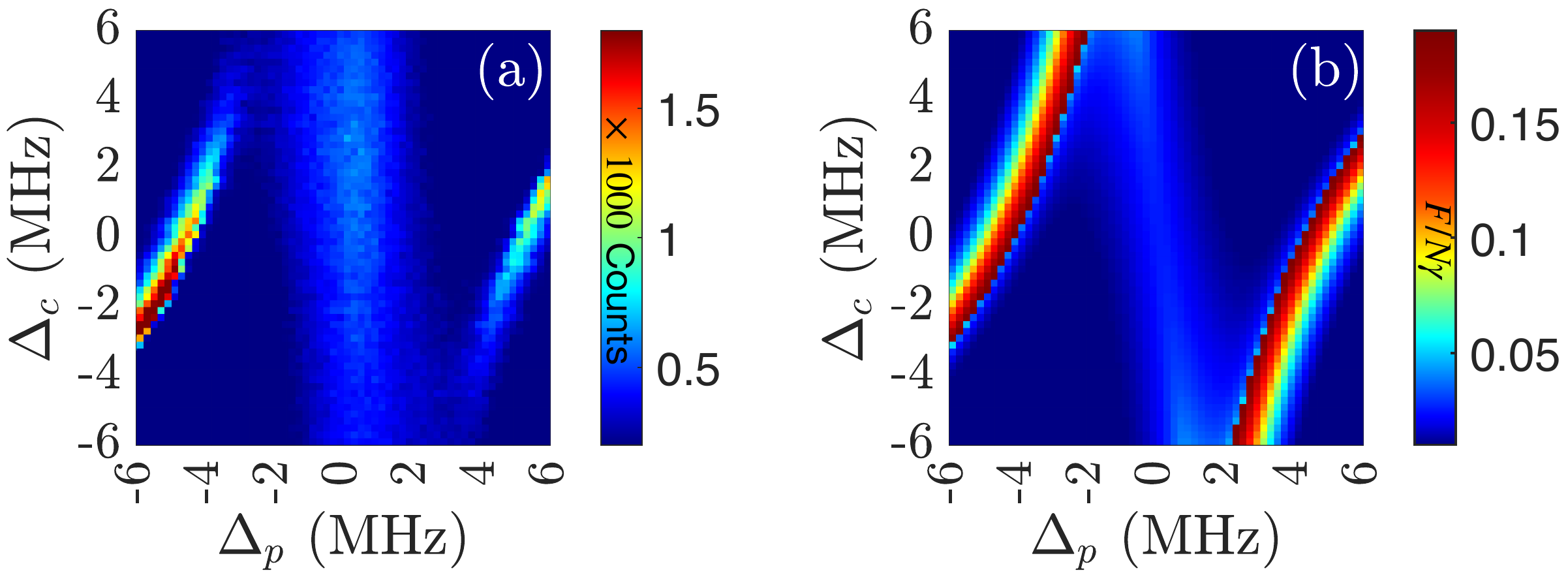}
    \caption{\label{Fig:4} Higher-resolution study of atomic fluorescence. (a) Measured and (b) calculated fluorescence are displayed vs.\ $\Delta_\text{c}$ and $\Delta_\text{p}$, exhibiting the atom-cavity coupled states alongside the broad resonance around $\Delta_\text{p}=0$. The parameters used for the calculation are same as in Fig.~\ref{Fig:Central}. A frequency resolution of \qty{200}{kHz} was used for both $\Delta_\text{c}$ and $\Delta_\text{p}$. }
\end{figure}

We extend our understanding by spanning the cavity-atom detuning over the same range as the probe-atom detuning, with the same resolution. Figure~\ref{Fig:4}(a) shows the measured fluorescence vs.\ $\Delta_\text{p}$ and $\Delta_\text{c}$. Figure~\ref{Fig:4}(b) shows the corresponding excited state population calculated from Eq.~\eqref{eq:F}. Apart from the coupled states as in Fig.~\ref{Fig:2}, the distinct peak around the atomic resonance is also visible. We find by exploring our theoretical model that its width increases with $\Delta\omega$, hence it is a consequence of the inhomogeneous broadening. It also requires significant driving strength, which indicates that it is a saturation feature. 
Remarkably, when introducing dephasing $\gamma_d\sim\Delta\omega$ we find that this central maximum flattens out and becomes invisible. Detecting the feature in our experiment therefore indicates that our inhomogeneous broadening exceeds the homogeneous dephasing rate. 

\section{Discussion and Conclusion\label{Sec:5}}

In this work we have investigated the formation of atom-cavity dressed states in the collective strong coupling regime on a narrow atomic line, while the atoms are trapped and cooled in a magneto-optical trap. The coupled states are observed in the cavity transmission as well as in free-space fluorescence, which enables us to simultaneously investigate their field character and their atomic character. A theoretical mean-field description reproduces the main features of the experimental observations. A broad atomic resonance which only appears in fluorescence is also found in the theory prediction, confirming our understanding. Hence it allows us to study how inhomogeneous broadening, by the much broader transition used for trapping and cooling, effects the atom-cavity interaction on the narrow line. 
 
Our simplified theory models effects such as frequency broadening arising from atomic motion, due to magnetic field gradients and fluctuations, and scattering of blue photons, by two simple mechanisms: inhomogeneous broadening introduced by $\Delta\omega$ and homogeneous dephasing described by $\gamma_\text{d}$. Despite of these simplifications we find satisfactory agreement between theory and experiment. In addition, it should be noted that the parameters used in the model calculations are consistent with other measurements on the same system \cite{Gothe:2019:2}. 

Nevertheless, we do not find full quantitative agreement between the theory and the experiment. The most striking difference is visible in the dressed state resonances which are far more pronounced in the theory than they are in the experimental measurement. 
Our theory predicts stable dressed states even for driving frequencies close to resonance, $\Delta_p\approx 0$. This is not found in the experiment and therefore requires refined modeling based on further measurements. One possibility might be a probe-frequency dependent source of dephasing and broadening which is particularly pronounced close to atomic resonance. Such a mechanism might be related to opto-mechanical forces coming from the green laser light. While such forces may be expected to have even significant effects \cite{Asboth:2005, Domokos:2002}, they have not been considered in the present paper but will be addressed in the future.

Our experiment and its modeling demonstrate an approach to study the formation and properties of atom-cavity states on a narrow line, far in the collective strong coupling regime, and when the atoms are cooled and trapped continuously on a much ($>$100-x) wider transition. Notably, we simultaneously investigate the field and the atomic character of the coupled states. Apart from the coupled states, we highlight a central resonance, visible only in the atomic signal, which allows us to draw conclusions regarding broadening and dephasing mechanisms in the atom-cavity system. Apart from fundamentally exploring collective latter-might phenomena, we expect our work to find significance, for example, in applications of narrow-line cavity coupling for ultra-stable lasers or active atomic clocks. 

\section*{Acknowledgments}
	We gratefully acknowledge funding by the Deutsche Forschungsgemeinschaft (DFG, German Research Foundation) through the Collaborative Research Center TRR-306 "QuCoLiMa", sub-project B04. S.B.J. acknowledges stimulating discussions with S. Schäffer and F. Fama and support from the Deutsche Forschungsgemeinschaft (DFG, German Research Foundation) through projects A4 and A5 in TRR-185 “OSCAR”.

\bibliography{references.bib}



\end{document}